\documentclass{article}

\usepackage{arxiv}

\usepackage[utf8]{inputenc} 
\usepackage[T1]{fontenc}    
\usepackage{hyperref}       
\usepackage{url}            
\usepackage{booktabs}       
\usepackage{amsfonts}       
\usepackage{nicefrac}       
\usepackage{microtype}      
\usepackage{cleveref}       
\usepackage{graphicx}
\usepackage{doi}
\usepackage{diagbox}

\title{Equivalence Test for Correlated Bivariate Binary Observation}

\author{Guanghui Huang \thanks{Corresponding Author.} \\
	Mathematics and Statistics College\\
	Chongqing University\\
	Chongqing,China,401331\\
	\texttt{hgh@cqu.edu.cn} \\
}

\renewcommand{\shorttitle}{Equivalence Test}

\hypersetup{
pdftitle={Equivalence Test},
pdfsubject={Statisitics},
pdfauthor={Guanghui Huang},
pdfkeywords={Bivariate Binary Distribution , Equivalence Test , Correlated Coefficient, Confidence Region, Size and Power}
}

\newtheorem{theorem}{Theorem}

\newtheorem{proposition}[theorem]{Proposition}

\begin{document}
	
\maketitle

\begin{abstract}
The bivariate binary model describes the population with two-dimensional attributes, where each dimension has two possible different values.
The equivalence test compares the performance of two competing methods through the difference between the two discordant probabilities. 
The observations of the two-dimensional variable are from the same sample, which makes it more possible that the two variables are mutually correlated.
This paper investigates the influence of the correlation between the two dimensions on the result of the equivalence test.
Under the null hypothesis, the marginal probability of the positive response is symmetric at any specified correlated coefficient,  and the discordance probability is also symmetric to the positive response probability. 
The marginal distribution function of the discordant observation is monotonically decreasing with the increase of the discordance probability.
And the minimum point of the distribution function is determined by the correlated coefficient.
Based on the joint distribution of the two discordant observations, a confidence region of the possible values of two discordant variables is proposed, which deduces an equivalence test with the marginal distribution of the discordance observation, called the margin test.
For a specified level of significance, the acceptance region of the McNemar test compares to the corresponding domain of the margin test for different sample sizes. 
The shape of the two kinds of acceptance regions is similar, except that the acceptance region of the margin test is slightly larger than the corresponding region of the McNemar test. 
The size and power of the McNemar test compare to the corresponding values of the margin test at a specified level of significance for different sample sizes. 
The risk of the type I error in both methods increases for a larger sample size or a smaller correlation coefficient, and the range of parameters where the margin test correctly accepts the null hypothesis is significantly larger than the corresponding range of the McNemar test.
Two real-world examples demonstrate how to understand the different decisions from the McNemar test and the margin test, where the observed data is on the boundary of the rejection regions. 
\end{abstract}


\section{Introduction}\label{introduction}


In biological and medical research, there is often a need to compare the accuracy of different measurements during diagnosis or the effectiveness of different therapeutic schedules during treatment. To quantitatively compare the differences between two or more methods, the non-inferiority test is used to determine whether the experimental index generated by one method differs from the same index generated by other methods beyond a specified level of magnitude. If the magnitude of the difference is chosen to be zero, the problem becomes a problem of checking whether the same experimental index generated by two or more methods is equivalent, which is called the equivalence test.\cite{liu_ma_wu_tai_2006,guo_luh_2020,sandie_molinari_etal_2022} used a non-inferiority test to compare the accuracy of continuous data from two measurements, and \cite{paul_tiwari_etal_2021,ghosh_guo_ghosh_2022} used non-inferiority test to compare the effectiveness of continuous data from three treatments.


Every sample has a two-dimensional attribute, and each dimension has two possible observations, such data can be described by the bivariate binary distribution, which yields the ordinary $2 \times 2$ contingent table \cite{book_Bishop_2007, fagerland_lydersen_laake_2013_same_unite}. 
For example, a patient receives two types of treatments successively, and the results of each treatment may be effective or ineffective. As the observations are from the same sample, there may be a correlation between the two treatment effects \cite{fay_lumbard_2020}. Another example is that the patients suffering from two different types of diseases are treated with the same drug, as the observed data come from two different groups, there may be a correlation between the group attribute of the samples and the drug efficacy\cite{wu_2018}. A third example is that patients with the same disease may have a positive or negative measurement result before receiving treatment, and the result of the same measurement method may also be positive or negative after treatment. Two times of measurement apply to the same sample, and there may be a correlation between the two measurement results \cite{fagerland_lydersen_laak_2013_sct}. 
It is important to accurately analyze the bivariate binary data to compare the accuracy of measurement schemes, evaluate the effect of therapeutic schedules, distinguish the application groups of new drugs, and evaluate the adverse effects of drugs. Some researchers assume that there is no correlation between the two dimensions \cite{liu_ma_wu_tai_2006,guo_luh_2020,sandie_molinari_etal_2022,ghosh_guo_ghosh_2022,rohmel_2005}, and this simplification may influence the decision of non-inferiority test or equivalence test \cite{wu_2018,kang_chung_ahn_2005,nam_kwon_2009, C_T_Le_1984}. Therefore, this paper focuses on the equivalence test of $2 \times 2$ contingent table when there is a correlated relationship between the two dimensions.


There are many results on the probability structure and parameter estimation method for the bivariate binary distribution. \cite{C_T_Le_1984} studied a symmetric distribution derived from the bivariate binary distribution, determined the probability structure, and gave a parameter estimation algorithm under the assumption that the two marginal positive probabilities are equivalent. \cite{albert_ingram_1985} discussed a series of distributions derived from the bivariate binary distribution, and pointed out that the $2 \times 2$  contingent table is equal to a four-dimensional multinomial distribution with four joint probabilities.  
\cite{manuel_weib_silva_2014,wu_rai_yan_2019} gave the upper bound and lower bound of the correlated coefficient using the marginal positive probabilities, and \cite{wu_rai_yan_2019} proposed a maximum likelihood estimation of the marginal probabilities and the correlated coefficient. Under the hypothesis of equivalence, this paper constructs a two-variable model for the parameter space using the correlated coefficient and the marginal positive probability, discusses the symmetric properties of the marginal positive probability and the discordant probability, establishes the monotonicity of the cumulative marginal distribution function of discordance observation, and determines the relationship between the minimum point of the cumulative marginal distribution function with the correlated coefficient.


Based on the exact or conditional distributions of the bivariate binary distribution, a series of unconditional or conditional p-value methods apply to the non-inferiority/equivalence test.
\cite{rohmel_2005} considered the constrained maximum likelihood estimation and the monotonicity of the proposed estimator. 
\cite{kang_chung_ahn_2005} constructed the conditional sample space and proposed the exact/conditional p-value methods by removing the redundant parameters.
\cite{nam_kwon_2009} compared the performances of the adjusted score method, a Wald-type test, and a modification of Obuchowski's method for clustered matched-pair data. And it is observed that sufficient sample size improves the accuracy of the equivalence test.
Using the monotonicity of Barnard convex set, \cite{sidik_2003} proposed two p-value tests and found that the probability of the type I error may exceed the specified nominal significance level for the tests derived from the asymptotic distribution. 
\cite{lloyd_moldovan_2011} considered the implementation order of three transforms and found that a proper order of combination improves the power of the p-value test.
\cite{li_liu_goldberg_2011} considered three kinds of p-value tests and calculated their exact size under the conditional monotonicity condition.
\cite{liu_hsueh_hsieh_chen_2002} compared the size and power of two asymptotic tests, a Wald-type test, and a restricted maximum likelihood estimation. 
\cite{fagerland_lydersen_laake_2014} found the best tests were the asymptotical McNemar test and the McNemar mid-p test among 24 tests and confidence intervals.
Based on the marginal distribution of the two discordance observations and the estimated confidence region, this paper presents an equivalence test called the margin test. And the influence of sample size, correlation coefficient, and nominal significance level on the size and power of the proposed test is discussed in detail.


The McNemar test is the most frequently used method in the equivalence test for bivariate binary distribution. \cite{wu_2018} studied an adjusted McNemar test and calculated its power with simulated data. \cite{fagerland_lydersen_laak_2013_sct} compared the performances of a mid-p test, an asymptotic McNemar test, an exact conditional McNemar test, and an exact unconditional test through their size and power. 
\cite{mohammadi_atashin_hofman_tan_2018} compared the accuracy of two ontology alignment systems over a matching task by the McNemar test. \cite{murtza_khan_akhtar_2019} compared the accuracy of several methods in target detection problems by the McNemar test. The wide application of the McNemar test in biomedical and other fields motivates us to discuss the size and power with different sample sizes, correlation coefficients, and significance levels in detail. 
We also compare the performances of the McNemar test and the proposed margin test on two real-world data sets, where the observed data is on the boundary of the rejection regions.


The rest of this paper is organized as follows. Section 2 introduces the bivariate binary model and its notations. 
The upper and lower bounds of the correlated coefficient are given in section 3. 
Section 4 depicts the parameter space of the null hypothesis with the positive probability and the correlated coefficient. And the symmetric property of the positive probability and the discordant probability is also discussed with the proposed parameter space.  
The monotonicity of the cumulative distribution function of the discordant observation is presented in section 4. And the relationship between the minimum point of the cumulative distribution function and the correlated coefficient is also discussed in section 4.
Section 5 gives a confidence region of the two discordance observations by their joint distribution. 
And an equivalence test is deduced from the proposed confidence region with the marginal distribution of the discordant observation. 
Section 6 briefly introduces the McNemar test and its rejection region.
The performance of the McNemar test and the margin test is compared through the range of acceptance region, size, and power in section 7. 
Section 8 compares the performance of the McNemar test and the margin test with two real-world examples.
The summary and discussion are given in section 9.

\section{Bivariate Binary Model}\label{bivariate_binary_model}

\begin{table}[h]
	\caption{The Bivariate Binary Distribution $X=(X_1, X_2)$.}
	\label{bivariate_binary_table_model_1}
	\centering
	\begin{tabular}{l|cc|c}
		\hline
		\diagbox{$X_1$}{$X_2$} & 0 & 1 & \\
		\hline
		0 & $p_{00}$ & $p_{01}$ & $p_{0+}$  \\
		1 & $p_{10}$ & $p_{11}$ & $p_{1+}$  \\
		\hline
		& $p_{+0}$ & $p_{+1}$ &   \\
		\hline
	\end{tabular}
\end{table}

The population $X=(X_1,X_2)$ follows the bivariate binary distribution as Table \ref{bivariate_binary_table_model_1} \cite{book_Bishop_2007}.
Denote the joint distribution of the two-dimensional variable $(X_1, X_2)$ as
\begin{equation}\label{prob:binary_joint}
	p_{ij}=Prob\{X_1=i,X_2=j\}, \quad i,j=0,1. 
\end{equation}
From the joint distribution (\ref{prob:binary_joint}), the marginal probabilities of $X_1$ and $X_2$ are 
\begin{eqnarray}
	&p_{0+} = p_{00}+p_{01},   \quad & p_{1+} = p_{10}+p_{11}, \\
	&p_{+0} = p_{00}+p_{10},   \quad &p_{+1}=p_{01}+p_{11},  \\
	&p_{0+} = 1 - p_{1+},      \quad &p_{+0} = 1-p_{+1}.
\end{eqnarray}
The marginal distributions of $X_1$ and $X_2$ are determined by the two positive probabilities $p_{1+}$ and $p_{+1}$ respectively. Denote the variance and standard deviation of $X_1$ and $X_2$ as
\begin{eqnarray}
	&\sigma^2_1 = D(X_1) =  p_{1+}(1-p_{1+}), \quad &\sigma_1 = \sqrt{\sigma_1^2}, \\ 
	&\sigma^2_2 = D(X_2) =  p_{+1}(1-p_{+1}), \quad &\sigma_2 = \sqrt{\sigma_2^2}.   
\end{eqnarray}
Denote the covariance of $X_1$ and $X_2$ as
\begin{equation}
	\theta = Cov(X_1,X_2) = E\left[ X_1X_2 \right] - E\left[X_1\right]E\left[X_2\right],
\end{equation}
and denote the correlated coefficient of $X_1$ and $X_2$ as
\begin{equation}\label{rho_exp}
	\rho = \frac{Cov(X_1,X_2)}{\sqrt{D(X_1)\cdot D(X_2)}}=\frac{p_{11}-p_{1+} p_{+1}}{\sqrt{p_{1+}(1-p_{1+}) p_{+1}(1-p_{+1}) }},
\end{equation}
it follows that
\begin{equation}
	\theta = \rho \sigma_1 \sigma_2.
\end{equation}

There are four parameters $ p_{00}$, $p_{01}$, $p_{10}$, $p_{11}$ in the bivariate binary distribution, with the constraint $p_{00} + p_{01} + p_{10} + p_{11} =1$, we need just three of them to describe the joint distribution.
In the following sections, another group of three parameters describes the bivariate binary distribution, including the two positive probabilities $p_{1+}$ and $p_{+1}$, and the correlated coefficient $\rho$.

\section{Bounds of Correlated Coefficient}

From the relationship between the correlated coefficient and the marginal positive probabilities (\ref{rho_exp}), we have
\begin{equation}
	p_{11} = p_{1+}p_{+1} + \rho  \sigma_1 \sigma_2,
\end{equation}
rewrite the two discordant probabilities as
\begin{eqnarray}
	p_{10} = p_{1+} - p_{11} = p_{1+} - p_{1+}p_{+1} - \rho \sigma_1 \sigma_2,  \\
	p_{01} = p_{+1} - p_{11} =  p_{+1}  - p_{1+}p_{+1} - \rho \sigma_1 \sigma_2. 
\end{eqnarray}
Notice that $p_{10} > 0 $ and $p_{01} > 0$, we have the following two inequalities 
\begin{eqnarray}
	&& \rho \sigma_1 \sigma_2 < p_{1+} - p_{1+}p_{+1},  \label{lower_1}\\
	&& \rho \sigma_1 \sigma_2 < p_{+1} - p_{1+}p_{+1},  \label{lower_2}
\end{eqnarray}
and the two inequalities (\ref{lower_1},\ref{lower_2})  should hold at the same time. If $p_{1+} > p_{+1}$ holds, the upper bound of $\rho$ satisfies
\begin{equation}
	\rho < \frac{p_{+1} - p_{1+}p_{+1}}{\sigma_1 \sigma_2} =  \sqrt{\frac{p_{+1}(1 - p_{1+})}{p_{1+}(1-p_{+1})}},
\end{equation}
and if $p_{1+} < p_{+1}$ holds, the upper bound of $\rho$ satisfies
\begin{equation}
	\rho < \frac{p_{1+} - p_{1+}p_{+1}}{\sigma_1 \sigma_2} = \sqrt{\frac{p_{1+}(1 - p_{+1})}{p_{+1}(1-p_{1+})}}.
\end{equation}

On the other hand, notice that $p_{10} < p_{1+}$ and $p_{01} < p_{0+} =  1 - p_{1+}$ should hold for the two discordant probabilities, therefore we have
\begin{eqnarray}
	&& p_{1+} - p_{1+}p_{+1} -\rho \sigma_1 \sigma_2  < p_{1+},\label{upper_1}\\	
	&& p_{+1} - p_{1+}p_{+1} - \rho \sigma_1 \sigma_2 < 1 - p_{1+}, \label{upper_2} 
\end{eqnarray}
where the two inequalities (\ref{upper_1},\ref{upper_2}) should hold at the same time. If $p_{1+} + p_{+1} - 1 > 0$ holds, the lower bound of $\rho$ satisfies
\begin{eqnarray}
	\rho \sigma_1 \sigma_2 & > &  p_{1+} + p_{+1} - 1 - p_{1+}p_{+1},   \\
	\rho    & > &  - \frac{(1 - p_{1+})(1 - p_{+1})}{\sqrt{p_{1+}p_{+1}(1-p_{1+})(1-p_{+1})}} \nonumber \\
	& = & - \sqrt{\frac{(1 - p_{1+})( 1 - p_{+1})}
		{p_{1+}p_{+1}}
	}.
\end{eqnarray}
On the other hand, if $p_{1+} + p_{+1} - 1 < 0$ holds, the lower bound of $\rho$ satisfies
\begin{eqnarray}
	\rho \sigma_1 \sigma_2 & > &   - p_{1+} p_{+1},   \\
	\rho & > & - \frac{p_{1+} p_{+1}} {\sqrt{p_{1+} p_{+1} (1 - p_{1+})( 1 - p_{+1})} } \nonumber \\
	& = &  - \sqrt{ \frac{p_{1+} p_{+1}}{(1 - p_{1+})(1 - p_{+1})} }.
\end{eqnarray}

The other constraints of the two discordant probabilities are $p_{10} < p_{+0} = 1 - p_{+1}$ and $p_{01} < p_{+1}$, which derive the same upper and lower bounds of $\rho$. Therefore we determine the upper and lower bounds of the correlated coefficient $\rho$ by the two positive probabilities $p_{1+}, p_{+1}$, which is the same as \cite{manuel_weib_silva_2014,wu_rai_yan_2019}.

\section{Parameter Space of the Null Hypothesis}

The purpose of the equivalence test is to determine whether the two positive probabilities $p_{1+}$ and $p_{+1}$ are equal to each other, which yields the following hypothesis test problem  
\begin{equation}
	H_{0}: p_{1+} = p_{+1}; \qquad H_{1}:   p_{1+} \neq p_{+1}.
\end{equation}
Denote $p_{1+} = p_{+1} = \pi$, we have
\begin{equation}
	\sigma_1^2 = p_{1+}(1-p_{1+}) = \pi(1-\pi), \quad \sigma_2^2 = p_{+1}(1-p_{+1}) = \pi(1-\pi) \Rightarrow \sigma_1^2 =\sigma_2^2. 
\end{equation}
When the null hypothesis holds, the two parameters $\pi$ and $\rho$ completely determine the parameter space.

\subsection{Bounds of Positive Probability}

When the null hypothesis $H_{0}$ holds, which means $p_{1+} = p_{+1} = \pi$, rewrite the upper bound of the correlated coefficient as 
\begin{equation}
	\rho < \sqrt{\frac{p_{+1}(1-p_{1+})}{p_{1+}(1-p_{+1})}} = \sqrt{\frac{p_{1+}(1-p_{+1})}{p_{+1}(1-p_{1+})}} = \sqrt{\frac{\pi(1-\pi)}{\pi(1-\pi)}} = 1,
\end{equation}
which imposes no additional constraint on the upper bound of $\rho$.

On the other hand, when $p_{1+}+p_{+1}-1>0$ holds, which means $\pi > 1/2$, rewrite the lower bound of
$\rho$ as
\begin{equation}
	\rho > \frac{p_{1+} + p_{+1} -1 -p_{1+}p_{+1}}{\sigma_1 \sigma_2} = -\frac{1-\pi}{\pi},
\end{equation}
which yields the upper bound of positive probability $\pi$ in terms of the correlated coefficient $\rho$ as
\begin{equation}
	\pi < \frac{1}{1-\rho},
\end{equation}
where $\pi > 1/2$ holds. 

When $p_{1+}+p_{+1}-1<0$ holds, which means $\pi < 1/2$, rewrite the lower bound of the correlated coefficient $\rho$ as
\begin{equation}
	\rho > -\frac{p_{1+}p_{+1}}{\sigma_1 \sigma_2} = -\frac{\pi}{1-\pi}, 
\end{equation}
which yields the lower bound of positive probability $\pi$ in terms of $\rho$ as 
\begin{equation}
	\pi > - \frac{\rho}{1-\rho},
\end{equation}
where $\pi<1/2$ holds.

On the other hand, when the correlated coefficient $\rho > 0 $, we always have
\begin{equation}
	- \frac{\rho}{1-\rho} <  0  <  1  < \frac{1}{1-\rho}.
\end{equation}
When $\rho > 0$ holds, the range of positive probability $\pi$ should be
\begin{equation}
	0 < \pi < 1.
\end{equation}

On the other hand, the following inequality (\ref{ineq_rho_boundary}) always holds for a negatively correlated coefficient,
\begin{equation}\label{ineq_rho_boundary}
	0 < -\frac{\rho}{1-\rho} < \frac{1}{1-\rho} < 1,
\end{equation}
which yields the range of positive probability $\pi$ as
\begin{equation}
	-\frac{\rho}{1-\rho} <  \pi < \frac{1}{1-\rho},
\end{equation}
when $\rho<0$ holds. Therefore, we have the following proposition.
\begin{proposition}	
	When the null hypothesis $H_0$ holds, the parameter space $\Theta_0(\rho, \pi)$ is determined by $\rho$ and $\pi$, which satisfies
	\begin{equation}\label{parameter_space_H_0}
		\Theta_0(\rho, \pi) 
		= 
		\left\{ 
		0< \pi < 1,   0 \leq \rho < 1
		\right\} 
		\cup 
		\left\{ 		     
		-\frac{\rho}{1-\rho} <  \pi < \frac{1}{1-\rho},  -1 < \rho < 0
		\right\}.
	\end{equation}
\end{proposition}

\subsection{Symmetry of Two Probabilities}

If the correlated coefficient $\rho \in (-1,0)$ holds, the summation of the upper and lower bounds of positive probability $\pi$ satisfies
\begin{equation}
	\frac{-\rho}{1 - \rho}  + \frac{1}{1 - \rho } = 1,
\end{equation}
which indicates that the center of the positive probability $\pi$ is at $\pi^* = 1/2$. On the other hand, the positive probability lies in $[0,1]$ when the correlated coefficient is positive, which indicates that the center of $\pi$ is also at $\pi^* = 1/2 $. Therefore we have the following proposition. 
\begin{proposition}
	When the null hypothesis $H_{0}$ holds, the center of the positive probability $\pi$ is $\pi^* = \frac{1}{2}$.
\end{proposition}


When the null hypothesis $H_{0}$ holds, the two positive probabilities satisfy $p_{1+} = p_{+1} = \pi $, such that the consistent probability $p_{11}$ satisfies
\begin{equation}
	p_{11} = p_{1+} p_{+1} + \rho \sigma_1 \sigma_2 = \pi^2 + \rho \pi(1 - \pi). 
\end{equation}
And the two discordant probabilities are equal to each other when the null hypothesis holds, 
which derives $p_{10} = p_{01} = \pi - p_{11}$, thus we have
\begin{equation}
	p_{10} = p_{01} = \pi - p_{11} = \pi -  \left( \pi^2 + \rho \pi (1 - \pi) \right) = \pi (1 - \pi ) (1 - \rho).   
\end{equation}
If $\rho < 0$ holds, $\pi$ reaches its minimum at the lower bound $\pi_{min} = -\frac{\rho}{1 - \rho}$, and reaches its maximum at the upper bound $\pi_{max} = \frac{1}{1 - \rho}$. Notice that the discordant probability satisfies
\begin{eqnarray}
	p_{10} (\pi_{min})= p_{01}(\pi_{min}) & = & -\frac{\rho}{1 - \rho} \left(1 + \frac{\rho}{1 - \rho}\right) (1 - \rho) \\
	&  =  &   -\frac{\rho}{1 - \rho}, 
\end{eqnarray}
and  
\begin{eqnarray} 
	p_{10} (\pi_{max})= p_{01}(\pi_{max})
	&  = & \frac{1}{1 - \rho} \left(1 - \frac{1}{1 - \rho}\right) (1 - \rho) \\
	&  =  &   -\frac{\rho}{1 - \rho} ,
\end{eqnarray}
which derives the following equations at $\pi_{min}$ and $\pi_{max}$ 
\begin{eqnarray}
	p_{10} (\pi_{max}) & = &  p_{10}(\pi_{min}), \\
	p_{01} (\pi_{max}) & = &  p_{01}(\pi_{min}).   
\end{eqnarray}

Notice that the function $f(\pi) = \pi (1 - \pi)(1 - \rho)$ is symmetric about $\pi^* = \frac{1}{2}$ for any specified $\rho$,  therefore the discordant probability is symmetric about the positive probability $\pi$ for all $ \rho \in (0,1)$. And the discordant probability $p_{10}(\pi)$ reaches its maximum value at $\pi^* = \frac{1}{2}$, we have 
\begin{equation}
	p_{10,max} = p_{01,max}= p_{10}\left(\pi^*=\frac{1}{2}\right) = \frac{1}{2} \left( 1 - \frac{1}{2} \right) \left(1 - \rho \right) = \frac{1 - \rho}{4}.
\end{equation}
Notice that the function $f(\rho) = \frac{1 - \rho}{4}$ decreases with increasing $\rho$, we have the following proposition.
\begin{proposition}
	When the null hypothesis $H_{0}$ holds, for any specified $\rho$, the discordant probability $p_{10}$ and $p_{01}$ are symmetric about $\pi^* = \frac{1}{2}$, and we have	
	\begin{eqnarray}
		p_{10} & = &  p_{01} = \pi (1 - \pi)(1 - \rho), \\
		p_{10,max} & = & p_{01,max} = \frac{1 - \rho}{4}.		
	\end{eqnarray}
	The range of discordant probability satisfies
	\begin{equation}\label{max_p_10}
		0 < p_{10} = p_{01}  \leq \frac{1}{2},
	\end{equation}
	where $p_{10}$ reaches its maximum value when the correlated coefficient $\rho = -1$.
\end{proposition}


\subsection{Monotonicity of Distribution Function}

The bivariate binary distribution can be viewed as the four terms multinomial distribution with observed frequency number $(N_{10},N_{01},N_{00},N_{11})$, where the discordant observations $N_{10}, N_{01}$ follow the binomial distribution with the successful probability $p_{10}$ and $p_{01}$ respectively. When $N_{10} = x_{10}$, the cumulative distribution function of $N_{10}$ is
\begin{equation}
	F(x_{10}; n, p_{10}) = Prob\left\{ N_{10} \leq x_{10} \right\} = \sum_{k=0}^{x_{10}} \frac{n!}{k!(n-k)!}p_{10}^{k}(1 - p_{10})^{n-k},
\end{equation}
where $n$ is the number of all four kinds of observations. For any fixed $x_{10}$ and $n$, the distribution function $F(x_{10};n, p_{10})$ is determined by $p_{10}$. Denote $x_{10} = x, p_{10}=p, C_n^k = \frac{n!}{k!(n-k)!}$ and $F(x_{10};n, p_{10}) = F(p)$, the first order derivation about $p$ is
\begin{eqnarray}\label{comulated_distribution_function}
	\frac{dF(p)}{dp} & = &  \sum_{k=0}^{x} C_n^k k p^{k-1} (1-p)^{n-k} - \sum_{k=0}^{x} C_{n}^k p^k (n-k)(1-p)^{n-k-1} \nonumber \\
	& = & \frac{1}{p} \sum_{k=0}^{x} k C_n^k p^{k} (1-p)^{n-k} - 
	\frac{1}{1-p}  \sum_{k=0}^{x} (n-k) C_n^k p^{k} (1-p)^{n-k}.
\end{eqnarray}
Denote the random index function of set $A=\{ k: k \leq x \}$ as
\begin{equation}
	I_{A}(N=k) = \left\{ 
	\begin{array}{ll}
		1, & k \in A= \{k: k \leq x \};\\
		0, & k \notin A.
	\end{array}
	\right.	
\end{equation}

On the other hand, when $N \sim B(n,p)$ holds, the conditional expectation of $N$ under the condition $\{N \leq x\}$ is
\begin{eqnarray}
	E[N | N \leq x ]     & = & \frac{1}{P\{A\}} E[N I_{A}(N)] = \frac{1}{P\{A\}} \sum_{k=0}^{n} k I_{A}(k) C_{n}^k p^k (1-p)^{n-k},   \\
	E[(n-N) | N \leq x ] & = & \frac{1}{P\{A\}} E[(n-N) I_{A}(N)] \nonumber \\
	& = & \frac{1}{P\{A\}} \sum_{k=0}^{n} (n-k) I_{A}(k) C_{n}^k p^k (1-p)^{n-k},
\end{eqnarray}
therefore (\ref{comulated_distribution_function}) can be expressed with conditional expectation as
\begin{eqnarray}
	\frac{dF(p)}{dp} &  = & \frac{P\{ A \}}{p} E[N | N \leq x] - \frac{P\{ A \}}{ 1 - p} E[(n-N) | N \leq x] \\
	& = & 
	\frac{ P\{A\} }{p} \left\{  E[N | N \leq x]  - \frac{p}{1-p} E[(n-N) | N \leq x] \right\} \\
	& = &
	\frac{P\{A\}}{p} \left\{  E[N | N \leq x]  - \frac{p}{1-p} n + \frac{p}{1-p} E[N | N \leq x] \right\} \\
	& = & 
	\frac{P\{A\}}{p(1-p)} \left\{  E[N | N \leq x]  - n p  \right\} \\
	& = & 
	\frac{P\{A\}}{p(1-p)} \left\{  E[N | N \leq x]  - E[N]  \right\} \\
	& = &
	\frac{ P\{A\} }{ p(1-p) }
	\left\{  E[N | N \leq x]  - P\{ N \leq x \}   E[N | N \leq x] \right.\nonumber \\
	& & \left. -   (1 - P\{N \leq x \} ) E[N | N > x ]  \right\} \\
	& = & 
	\frac{ P\{ A\} ( 1 - P\{N \leq x \} ) }{ p(1-p) } \left\{ E[N | N \leq x] - E[N | N > x]  \right\}. 	   \label{two_condition_probs}
\end{eqnarray}	
Notice that
\begin{eqnarray}
	E[N | N \leq x] & \leq & x, \\
	E[N | N > x ]   & \geq    & x+1,
\end{eqnarray}	
subtract to (\ref{two_condition_probs}), we have
\begin{equation}
	\frac{dF(p)}{dp} \leq   \frac{ P\{ A\} ( 1 - P\{ A \} ) }{ p(1-p) } \left[ x - (x+1) \right] \leq   - \frac{ P\{ A\} ( 1 - P\{A\} ) }{ p(1-p) } \leq 0,
\end{equation}
and the following proposition holds.
\begin{proposition}
	For any fixed number of trails $n$ and the observed frequency $x$, the cumulative distribution function of binomial distribution $F(x;n,p)$ decreases with increase of probability $p$, e.g. if $ p_{1} < p_{2} $, we have 
	\begin{equation}
		F(x;n, p_{1}) \geq F(x; n, p_{2}).
	\end{equation}
\end{proposition}

Notice (\ref{max_p_10}), the distribution function $F(x_{10};n,p_{10})$ of $N_{10}$ satisfies the following proposition.
\begin{proposition}
	When the null hypothesis $H_{0}$ holds, for any $-1< \rho <1$ and $0< p_{10} < 1$, it follows 
	\begin{equation}
		F(x_{10};n,p_{10}) \geq F(x_{10};n,\frac{1-\rho}{4}),
	\end{equation}
	which means that for any fixed $x_{10}$ and $n$, we have 
	\begin{equation}
		\min_{p_{10}}F(x_{10};n,p_{10}) = F(x_{10};n,\frac{1}{2}).
	\end{equation}
\end{proposition}
The previous results state that if we consider the influence of the correlated coefficient $\rho$ between the two variables $(X_{1},X_{2})$ on the distribution function of discordant observation $N_{10}$, the distribution function $F(x_{10};n,p_{10})$ will reach its minimum value at the point $p_{10} = \frac{1}{2}$, which actually assumes that $\rho = -1$ holds.


\section{Confidence Region}

For any fixed sample size $n$ and discordant probabilities $p_{10}, p_{01}$, denote the observed discordant numbers $(N_{10}, N_{01})$ as $(x_{10}, x_{01})$, the sample space of the two discordant observations is
\begin{equation}
	\Omega = \left\{ (x_{10}, x_{01}) : 0 \leq x_{10} + x_{01} \leq n, 0 \leq x_{10}, x_{01} \leq n \right\}.
\end{equation}
Denote the significiant level as $1 - \alpha $, and the corresponding confidence region $A_{\alpha}$ as
\begin{equation} \label{trust_interval}
	A_{\alpha} = \left\{ (x_{10} , x_{01}): 
	L_{10} \leq x_{10} \leq U_{10},
	L_{01}(x_{10}) \leq x_{01} \leq  U_{01}(x_{10})  \right\}, 
\end{equation}
where $L_{10},U_{10},L_{01}(x_{10}),U_{01}(x_{10})$ denote the corresponding lower and upper bounds respectively, thus the probability of $\left\{ \left(N_{10},N_{01}\right) \in A_{\alpha} \right\}$ is
\begin{eqnarray}
	& & Prob\left\{ \left(N_{10}, N_{01}\right) \in A_{\alpha} \right\} 
	= 
	\sum_{n_{10} = L_{10}}^{U_{10}}
	\sum_{n_{01} = L_{01}(n_{10})}^{U_{01}(n_{10})} Prob\left\{N_{10}=n_{10}, N_{01} = n_{01}\right\} \\
	& & = 
	\sum_{n_{10} = L_{10}}^{U_{10}}
	\sum_{n_{01} = L_{01}(n_{10})}^{U_{01}(n_{10})} Prob\left\{N_{10}=n_{10}\right\} 
	Prob\left\{N_{01} = n_{01} | N_{10}=n_{10} \right\}   \\
	& & = 
	\sum_{n_{10} = L_{10}}^{U_{10}}  Prob\left\{N_{10}=n_{10}\right\} 
	\sum_{n_{01} = L_{01}(n_{10})}^{U_{01}(n_{10})}
	Prob\left\{N_{01} = n_{01} | N_{10}=n_{10} \right\} \label{two_comb_prob}.	  
\end{eqnarray}
If the inner summation of the conditional probabilities is a constant  $\omega$, i.e.
\begin{equation}
	\sum_{n_{01} = L_{01}(n_{10})}^{U_{01}(n_{10})}
	Prob\left\{N_{01} = n_{01} | N_{10}=n_{10} \right\}  = \omega, 
\end{equation}
and the outer summation of the unconditional probabilities is the same constant $\omega$, i.e.
\begin{equation}
	\sum_{n_{10} = L_{10}}^{U_{10}}  Prob\left\{N_{10}=n_{10}\right\}  = \omega,
\end{equation}
which makes (\ref{two_comb_prob}) can be rewritten as
\begin{equation}
	Prob\left\{ \left( N_{10}, N_{01} \right) \in A_{\alpha} \right\} 
	= \omega^2 = 1 - \alpha,
\end{equation}
where  $\alpha$ is the significant level of the confidence region. Therefore we have
\begin{equation}
	\sum_{n_{10} = L_{10}}^{U_{10}}  Prob\left\{N_{10}=n_{10}\right\}  = \omega =\sqrt{1 - \alpha},
\end{equation}
and 
\begin{equation}
	\sum_{n_{01} = L_{01}(n_{10})}^{U_{01}(n_{10})}
	Prob\left\{N_{01} = n_{01} | N_{10}=n_{10} \right\}  = \omega =  \sqrt{1 - \alpha}. 
\end{equation}

To determine the acceptance region $A_{\alpha}$, we just need to determine the lower bound $L_{10}$ and upper bound  $U_{10}$ for the first discordant variable $N_{10}$, and to determine the corresponding conditional upper and lower bound  $U_{01}(n_{10})$ and $L_{01}(n_{10})$ of the second discordant variable $N_{01}$, which satisfy the following constraints 
\begin{eqnarray}
	L_{10} &=& \arg\max_{L} Prob\left\{ N_{10} < L \right\}  \leq   \frac{1- \sqrt{1 - \alpha}}{2}, \label{margin_test_1}\\
	U_{10} &=& \arg\min_{U} Prob\left\{ N_{10} < U \right\}  \geq  1 - \frac{1- \sqrt{1 - \alpha}}{2}, \label{margin_test_2}\\
	L_{01}(n_{10}) &=& \arg\max_{L}	Prob\left\{ N_{01} < L  | N_{10} = n_{10}\right\}  \leq  \frac{1- \sqrt{1 - \alpha}}{2}, \label{margin_test_3}\\
	U_{01}(n_{10}) &=& \arg\min_{U}	Prob\left\{ N_{01} < U | N_{10} = n_{10}\right\}  \geq   1 - \frac{1- \sqrt{1 - \alpha}}{2}. \label{margin_test_4} 
\end{eqnarray}
We construct the confidence region $A_{\alpha}$ of the two-dimensional variable $(N_{10},N_{01})$
through their marginal distribution and conditional distribution. 

When the null hypothesis $H_0$ holds,  i.e. $p_{10} = p_{01} = p$, the maximum likelihood estimation of the discordant probability $p_{10}$ and $p_{01}$ can choose to be 
\begin{equation}
	\hat{p}_{10} = \hat{p}_{01}= \frac{x_{10} + x_{01}}{2n}.
\end{equation}

The frequency $N_{10}$ and $N_{01}$ both follow the binomial distribution $B(n,p)$ respectively. For a fixed sample size $n$ and significant level $\alpha$, denote the left and right endpoints
as $L_{\alpha}$ and $U_{\alpha}$ respectively, which satisfy the following two conditions
\begin{eqnarray}
	& L_{\alpha}   &= \arg\max_{L}Prob\{ N < L \} \leq \frac{1 - \sqrt{1 - \alpha}}{2}, \\
	& U_{\alpha}   &= \arg\min_{U}Prob\{ N < U \} \geq  1 - \frac{1 - \sqrt{1 - \alpha}}{2}, 
\end{eqnarray}
where $N \sim B(n,\hat{p}_{10})$. Therefore the rejection region for the null hypothesis $H_{0}$ can be 
\begin{equation}\label{rejection_margin_test}
	R_{\alpha} = \left\{ \min\left\{x_{10}, x_{01} \right\} < L_{\alpha} \right\} \cup  \left\{\max\left\{x_{10}, x_{01} \right\} > U_{\alpha} \right\}.
\end{equation}
Because the marginal distribution of the two discordant observations $(N_{10}, N_{01})$ is used to construct the rejection region, the proposed method is called the margin test with significant level $\alpha$ thereinafter.


\section{McNemar test}

The most frequently used equivalence test method is the McNemar test \cite{McNemar_test_1947}, and the statistic is 
\begin{equation}\label{McNemar}
	M = \frac{\left(x_{10} - x_{01}\right)^2}{x_{10} + x_{01}} \sim \chi^2(1). 
\end{equation}
The corresponding rejection region of McNemar test is
\begin{eqnarray}\label{McNemar_rejection}
	R_{\alpha} & = &  \left\{ M > \chi^2_{1-\alpha}(1)\right\}, \\
	& = & \left\{\left(x_{10},x_{01}\right) \in \Omega:  \frac{\left(x_{10} - x_{01}\right)^2}{x_{10} + x_{01}} > \chi^2_{1 - \alpha}(1)   \right\},
\end{eqnarray}
where $\chi^2_{1 - \alpha}(1)$ is the $1 - \alpha$ left quantile of the $\chi^2$-distribution with $1$  degree of freedom.
Although there are many improved versions of the McNemar test, the traditional version (\ref{McNemar}) is considered in this paper, as the focus of this paper is to investigate the influence of the correlated coefficient on the result of the equivalence test.


\section{Evaluation of Tests}

The size and power are the technical indexes to evaluate the performance of the equivalence tests in this paper. When the null hypothesis $H_0$ holds, the two positive probabilities equal to each other, i.e. $p_{1+} = p_{+1} = \pi$, and the possible range of parameter value lies in the space $\Theta_0$, which is determined in (\ref{parameter_space_H_0}). Denote the parameter space of $H_{1}$ as 
\begin{equation}\label{parameter_space_H_1}
	\Theta_{1} = \left\{ (p_{10}, p_{01}) :  0 < p_{10} + p_{01} < 1, 0 < p_{10}, p_{01} < 1, p_{10} \neq p_{01} \right\}.
\end{equation}

Denote the rejection region as $R_{\alpha}$, and the corresponding acceptance region is $A_{\alpha}$. The size of test is the probability of making the type I error. For a pair of parameter $(\rho, \pi)\in \Theta_{0}$, the corresponding size is
\begin{equation}\label{size_definition}
	size(\rho,\pi) = Prob\{ (N_{10}, N_{01}) \in R_{\alpha} | (\rho, \pi) \in \Theta_0 \}.
\end{equation}
For any specified rejection region $R_{\alpha}$, (\ref{size_definition}) determines the value of size for each point $(\rho,\pi)$ in the parameter space $\Theta_0$, which yields the surface of size on the parameter space $\Theta_0$.  Taking the advantage of the two-variable model for the parameter space $\Theta_0$, it is possible to intuitively evaluate the performance of the test method through the explicit surface of size in the following sections. The surface of size directly shows the influence of the correlated coefficient on the result of the equivalence test.

On the other hand, the power of test method is the probability of correctly rejecting the null hypothesis for the parameters in the parameter space $\Theta_{1}$, i.e. 
\begin{equation}\label{power_definition}
	power(p_{10},p_{01}) = Prob\{ (N_{10}, N_{01}) \in R_{\alpha} | (p_{10}, p_{01}) \in \Theta_{1}\}.
\end{equation}
For a specified rejection region $R_{\alpha}$, calculate the probability following (\ref{power_definition}) for each point $(p_{10},p_{01})$ in the parameter space of the alternative hypothesis $H_1$, which yields the surface of power on the set $\Theta_{1}$.


\subsection{Acceptance Region and Rejection Region}

For a fixed sample size $n$, the McNemar test and the margin test can determine whether a point $\left(x_{10},x_{01}\right)$ in the sample space $\Omega$ belongs to the acceptance region $A_{\alpha }$, or the region of rejection $R_{\alpha}$ respectively. The sample size determines the range of sample space $\Omega$, and the level of significance determines the shape of the acceptance region.

\begin{figure}[h!]
	\begin{center}
		\includegraphics[width = 1.00 \textwidth]{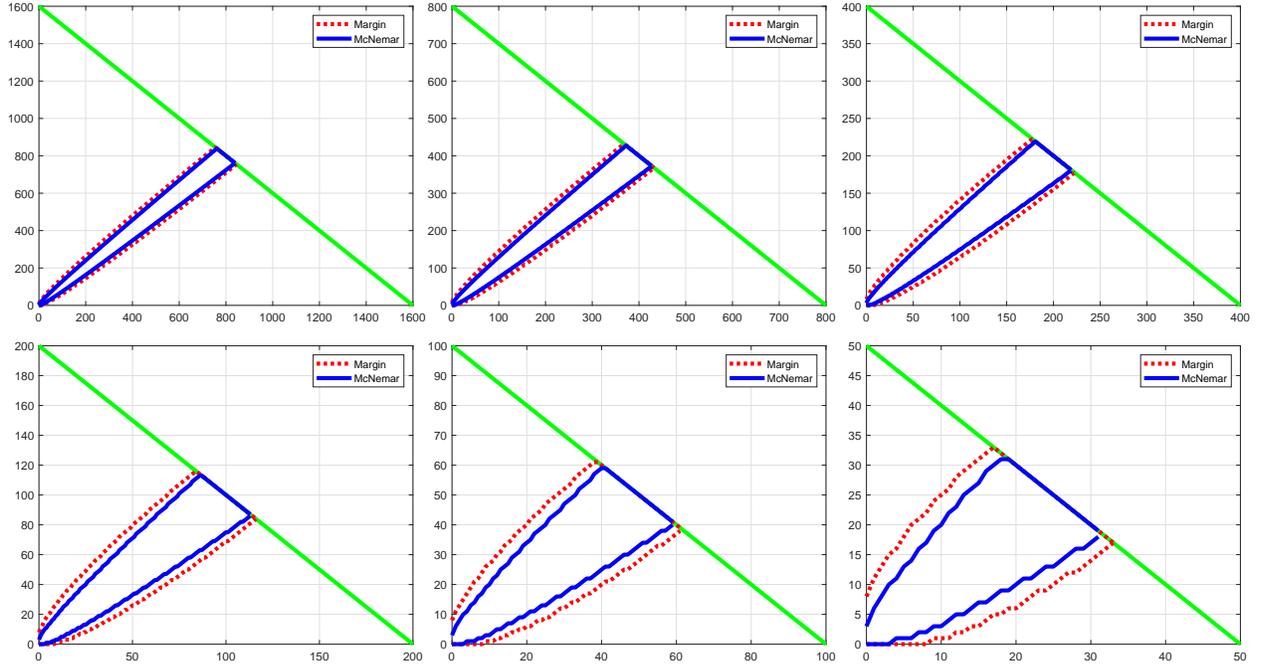}
		\caption{Boundaries of Acceptance Regions for $\alpha = 0.05, n=1600, 800, 400, 200, 100, 50$.}
		\label{fig:A_alpha}
	\end{center}	
\end{figure}

Under the significant level $\alpha = 0.05$, Fig.\ref{fig:A_alpha} shows the boundaries of acceptance regions of the McNemar test and the margin test for sample sizes $n = 1600,800,400,200,100,50$. The dotted red lines are the boundaries of the McNemar test, and the solid blue lines are the boundaries of the margin test. The shape of the two acceptance regions is similar, and the acceptance region of the margin test always wraps the corresponding domain of the McNemar test. Therefore the probability that a sample follows in the acceptance region of the margin test is slightly larger than the corresponding probability of the McNemar test.


\subsection{Size of Tests}

Denote the acceptance region of $H_{0}$ as
\begin{equation}
	A_{\alpha} = \left\{ (x_{10}, x_{01}): L_{10} \leq x_{10} \leq U_{10}, 
	L_{01}(x_{10}) \leq x_{01} \leq U_{01}(x_{10}) \right\},
\end{equation}
for any specified value of parameter $(\rho, \pi) \in \Theta_{0}$, the probability that a sample $(N_{10},N_{01})$ lies in the rejection region $R_{\alpha}$ is 
\begin{eqnarray}
	size(\rho,\pi) & = &  Prob\{ (N_{10}, N_{01}) \in R_{\alpha} | (\rho,\pi) \in \Theta_{0}\} \nonumber \\ 
	& = & 1 - Prob\{ (N_{10}, N_{01}) \in A_{\alpha} | (\rho,\pi) \in \Theta_{0}\} \nonumber \\
	& = & 1 - \sum_{n_{10}=L_{10}}^{U_{10}} Prob\left\{ N_{10} = n_{10} \right\}
	\sum_{n_{01}=L_{01}(n_{10})}^{U_{01}(n_{10})} Prob\left\{ N_{01} = n_{01} | N_{10} = n_{10} \right\},
\end{eqnarray}
where $N_{10} \sim B(n,p_{10})$,  $N_{01} | N_{10} = n_{10} \sim B(n - n_{10}, q_{01})$, and the conditional probability is
\begin{equation}
	q_{01} = \frac{p_{01}} {1 - p_{10}}.
\end{equation}

If the resulting size is below the specified value of significant level $\alpha$, the target of controlling the type I error is achieved. Otherwise, the value of size exceeds the significant level, which means the test does not satisfy the desired target of risk control.

\begin{figure}[h!]
	\begin{center}
		\includegraphics[width = 1.00 \textwidth]{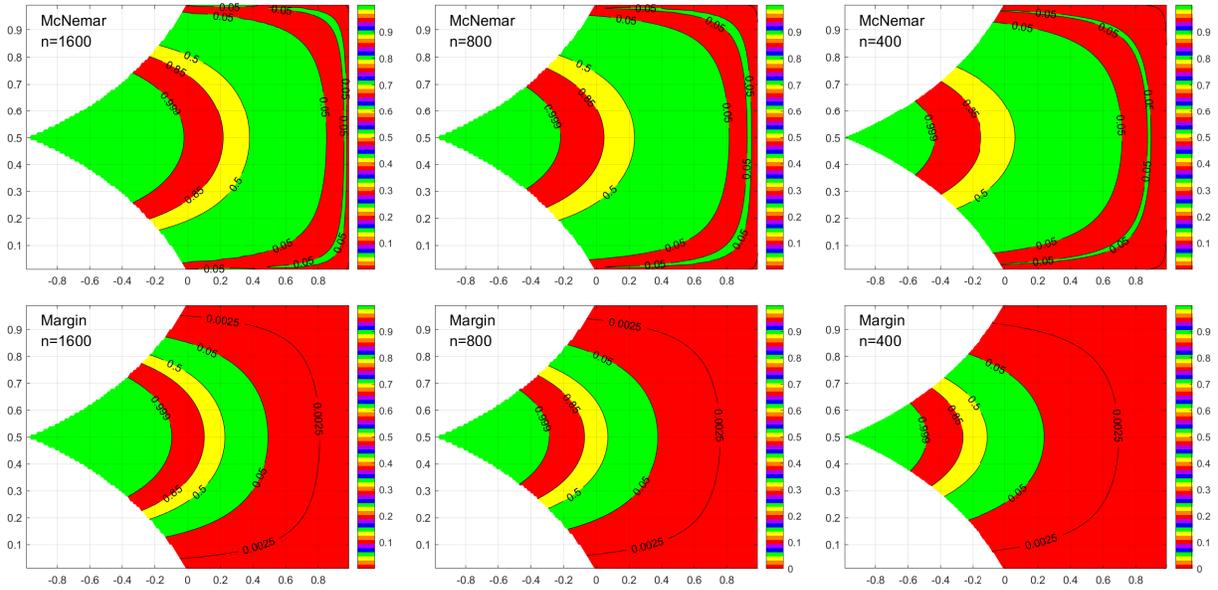}
		\caption{Contours of the Size Surfaces for $\alpha = 0.05, n=1600,800,400$.}
		\label{fig:size_large_n}
	\end{center}	
\end{figure}

\begin{figure}[h!]
	\begin{center}
		\includegraphics[width = 1.00 \textwidth]{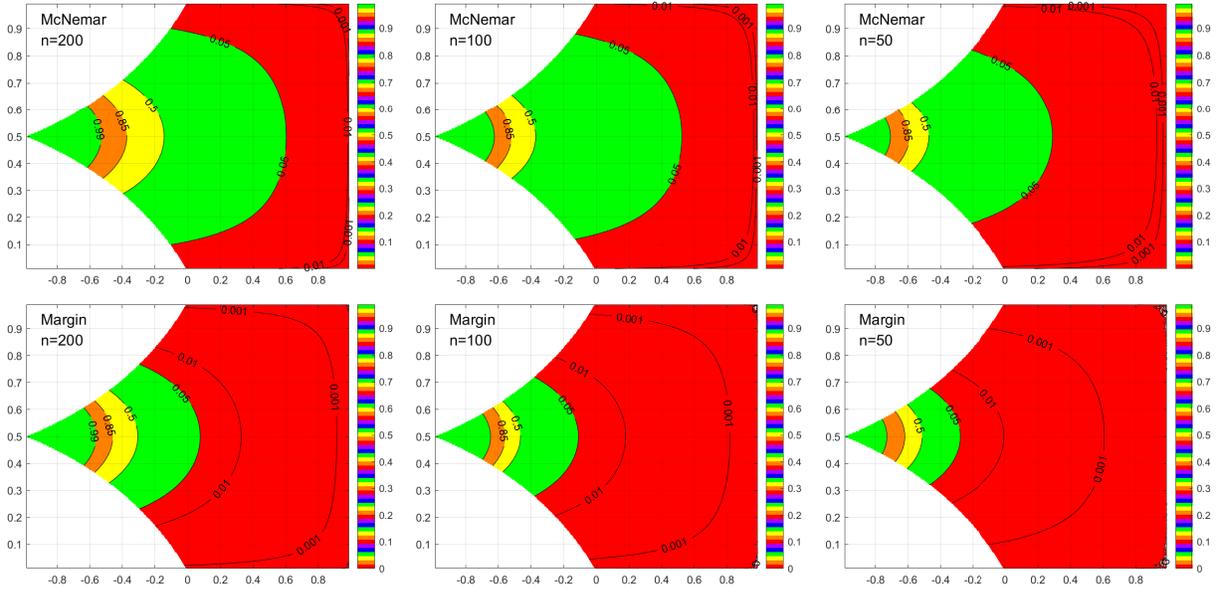}
		\caption{Contours of the Size Surfaces for $\alpha = 0.05, n=200, 100, 50$.}
		\label{fig:size_small_n}
	\end{center}	
\end{figure}

A two-variable model describes the parameter space of $H_0$ in this paper, which makes it possible to see the influence of the correlated coefficient on the value of size. Fig. \ref{fig:size_large_n} and Fig. \ref{fig:size_small_n} show the contours of size surfaces, where the horizontal coordinate indicates the value of correlated coefficient $\rho$, and the vertical coordinate indicates the value of positive probability $\pi$. 
When the correlated coefficient changes from $-0.99$ to $0.99$, the type I error shows a downward tendency, and the size is symmetric about $\pi^* = 1/2$. Note that the significant level $\alpha$ is $0.05$, those points covered by the contours below $0.05$ are the correctly accepted parameters. 
The acceptance region of the margin test wraps up the acceptance domain of the McNemar test, such that the range of correctly identifying parameters by the margin test is larger than the corresponding region of the McNemar test.

For the regions of positively correlated coefficient, the margin test can correctly identify almost all of the parameters in $\Theta_0$, while the McNemar test can only correctly identify a part of the parameters when the sample size $n$ is $100$ or $50$. For the negative correlation coefficients, the two methods can't correctly identify all of the parameters in $\Theta_0$, and the correctly identified region of the margin test is obviously larger than the corresponding domain of the McNemar test. From the Fig.\ref{fig:size_large_n} and Fig. \ref{fig:size_small_n}, we can also find that the size of both tests gradually increases with the increment of sample size, and the risk of the type I error for the McNemar test increases significantly than the increment of the margin test.

We can also find that the surface of size given by the McNemar test fluctuates up and down with the increase of correlation coefficient in three experiments with sample sizes $n=1600,800,400$. The contour line of $0.05$ repeatedly appears three times on the right side of the graph, which indicates that the risk of type I error for the McNemar test fluctuates around $\alpha$ with relatively larger $\rho$. 
We also find that the fluctuation of the size surface with the increase of correlation coefficient also appears for a relatively small sample size with a large value of $\alpha$. 
However, the surface of size given by the margin test has no such obvious fluctuation, which means that the probability of making the type I error decreases monotonously with the increase of the correlation coefficient. It can conclude that the margin test is more stable for different sample sizes and significance levels.


\subsection{Power of Tests}

For any specified rejection region $R_{\alpha}$ and acceptance region $A_{\alpha} = \Omega - R_{\alpha}$, denote the power of each parameter point $(p_{10},p_{01}) \in \Theta_{1}$ as 
\begin{eqnarray}
	& & power(p_{10},p_{01})  =  Prob\{ (N_{10}, N_{01}) \in R_{\alpha} | (p_{10},p_{01}) \in \Theta_{1} \} \nonumber \\
	& & =  1 -  Prob\{ (N_{10}, N_{01}) \in A_{\alpha} | (p_{10},p_{01}) \in \Theta_{1} \}  \nonumber \\
	&  & =   1 - \sum_{n_{10} = L_{10}}^{U_{10}} Prob\{N_{10} = n_{10}\}
	\sum_{n_{01} = L_{01}(n_{10})}^{U_{01}(n_{10})}  Prob\{ N_{01} = n_{01} | N_{10} = n_{10}\},
\end{eqnarray}
where the conditional distribution of $N_{01}$ under the condition $N_{10} = n_{10}$ is a binomial distribution with
\begin{equation}
	N_{01} | N_{10} = n_{10} \sim B(n - n_{10}, q_{01}),
\end{equation}
where the number of trails is $n-n_{10}$, and the successful probability $q_{01}$ satisfies
\begin{equation}
	q_{01} = \frac{p_{01}}{1 - p_{10}}.	
\end{equation}

\begin{figure}[h!]
	\begin{center}
		\includegraphics[width = 1.00 \textwidth]{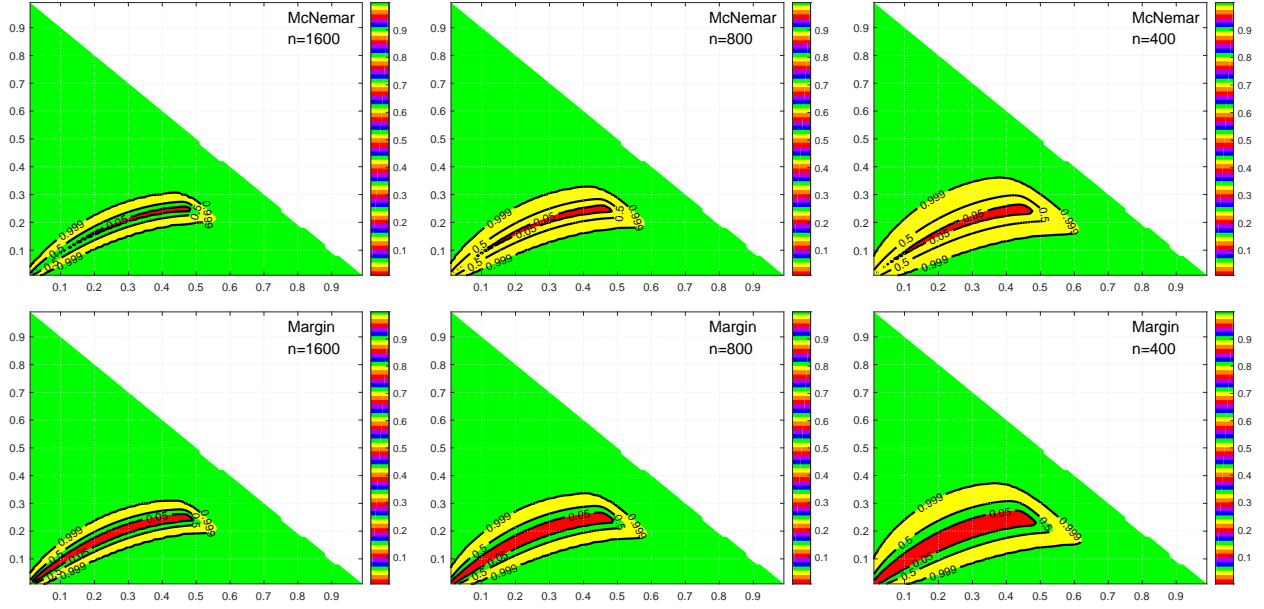}
		\caption{Contours of Power Surfaces with $\alpha = 0.05, n=1600,800,400$.}
		\label{fig:power_large_n}
	\end{center}	
\end{figure}

\begin{figure}[h!]
	\begin{center}
		\includegraphics[width = 1.00 \textwidth]{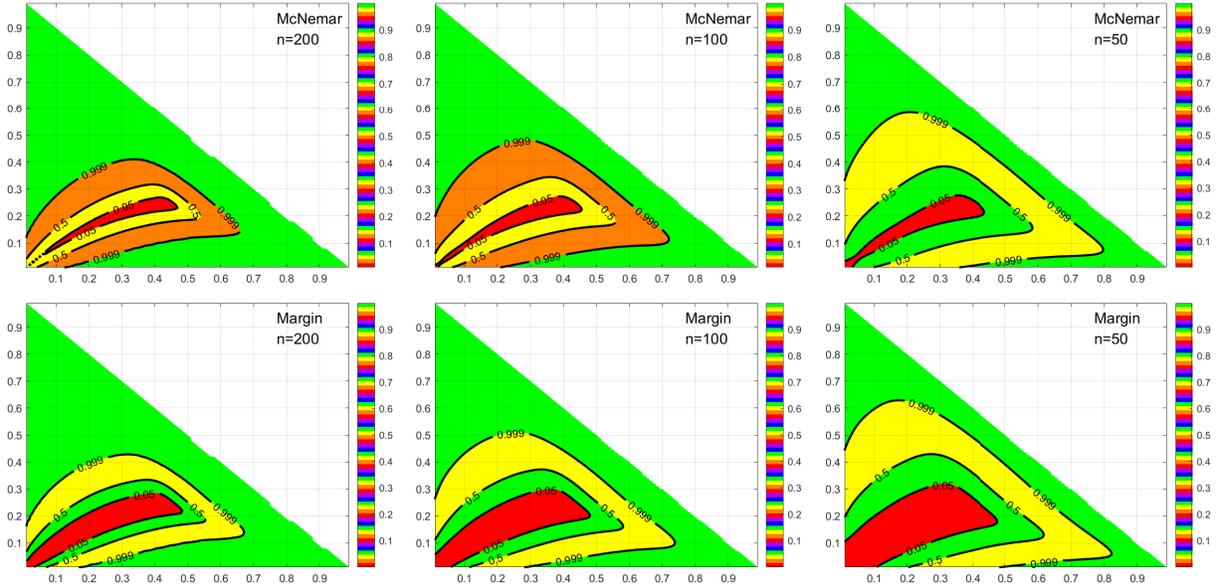}
		\caption{Contours of Power Surfaces with $\alpha = 0.05, n=200,100,50$.}
		\label{fig:power_small_n}
	\end{center}	
\end{figure}

\begin{figure}[h!]
	\begin{center}
		\includegraphics[width = 1.00 \textwidth]{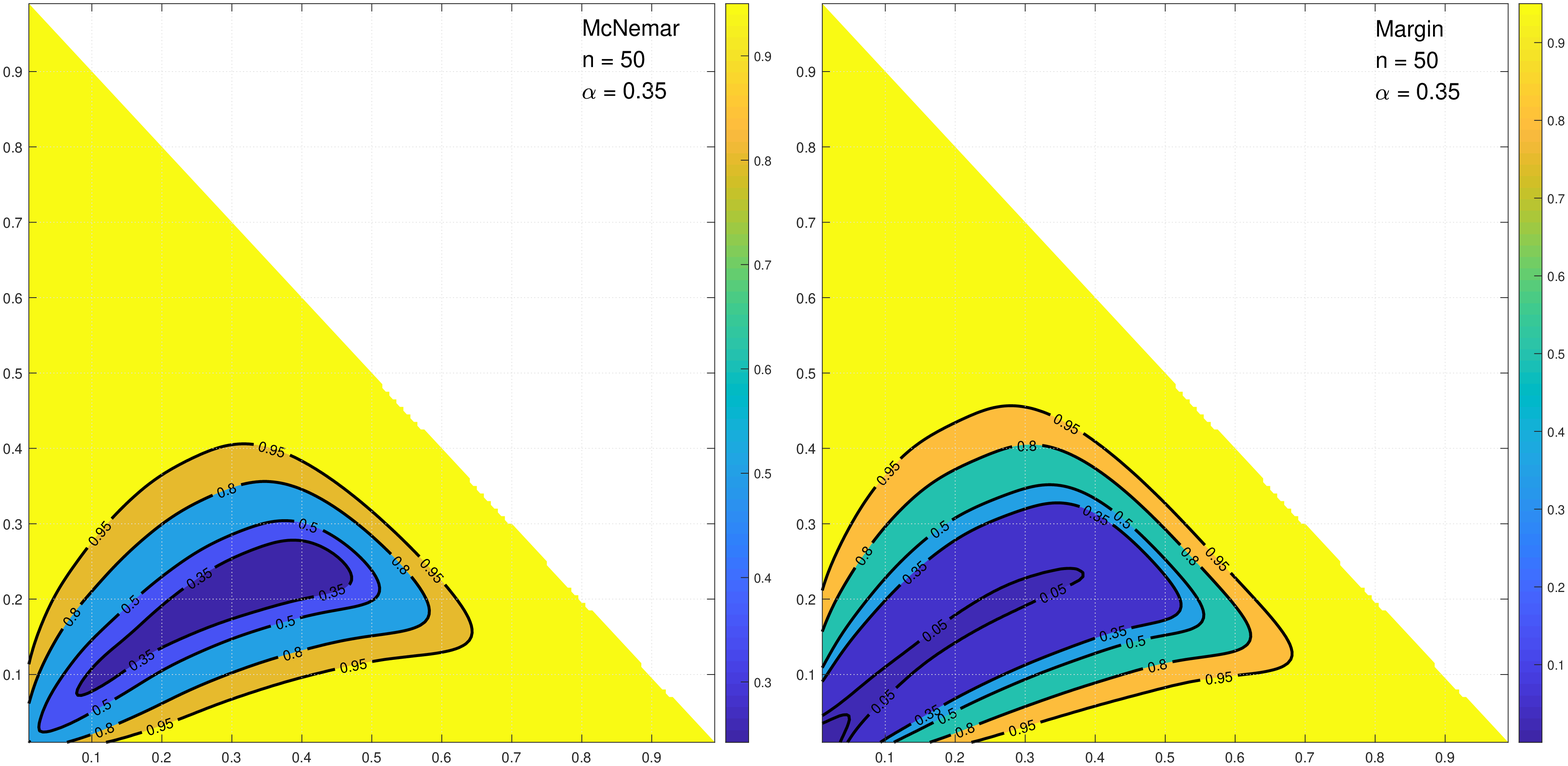}
		\caption{Contours of Two Power Surfaces with $\alpha = 0.35, n = 50$.}
		\label{fig:power_n_50}
	\end{center}	
\end{figure}

The contours of power surfaces of the McNemar test and the margin test are given in Fig.\ref{fig:power_large_n} and Fig.\ref{fig:power_small_n}, where the level of significant is $0.05$, and the samples sizes $n = 1600,800,400,200,100,50$. Another figure of the contours of two power surfaces is also given in Fig.\ref{fig:power_n_50}  for $\alpha = 0.35, n=50$. The horizontal coordinate indicates the value of $p_{10}$, and the vertical coordinate indicates the value of $p_{01}$ for the parameter $(p_{10},p_{01}) \in \Theta_{1}$.

From Fig. \ref{fig:power_large_n} and Fig. \ref{fig:power_small_n}, we can find that the region where the McNemar test and the margin test correctly reject the null hypothesis for the parameters in $\Theta_1$ gradually increases with the increase of sample size, and the region where the two tests accept the null hypothesis always concentrates around the diagonal line direction.

On the other hand, the probability of correctly accepting $H_0$ by the McNemar test tends to decrease with the reduction of $p_{10}$ and $p_{01}$. To observe the tendency of reducing, we give the Fig. \ref{fig:power_n_50}, where the significant level is relatively large, $\alpha = 0.35$, and the sample size is relatively small, $n = 50$. We can find that the region enclosed by the contour of $power = 0.35$ concentrates around the diagonal line. And the area given by the McNemar test separates from the original point, while the region given by the margin test closely connects to the original point. The observed connection shows that the margin test tends to correctly accept the null hypothesis for small $p_{10}$ and $p_{01}$, while the McNemar test tends to incorrectly reject the null hypothesis for small discordant probabilities.


\section{Real-World Examples}

To compare the performance of the McNemar test and the margin test with real-world data, especially for small sample sizes and those observed data on the boundary of the acceptance region, we discuss the following two real-world examples.

\subsection{Stem Cell Transplantation and Airway Hypertension}

In the first example, we discuss a small sample case. Twenty-one children underwent airway hypertension testing before and after stem cell transplantation (SCT) \cite{fagerland_lydersen_laak_2013_sct,bentur_lapidot_livnat_2009_SCT_example}, the observed data are shown in Table \ref{tab:21_SCT},
where $n = 21, n_{10} = 7, n_{01} = 1$, AHR means positive response, and NO AHR means negative response.

\begin{table}
	\begin{center}
		\caption{Airway Hypertension Test for 21 Children before and after Stem Cell Transplantation.}
		\label{tab:21_SCT}
		\begin{tabular}{ccccc}
			\toprule		
			&  &\multicolumn{2}{c}{After SCT} &  \\
			\cline{3-4}
			&  & AHR    & NO AHR & SUM \\
			\hline
			& AHR    &1  & 1     &2 \\
			Before SCT	& NO AHR            &7  & 12     & 19 \\	      	
			& SUM              &8 & 13   & 21 \\		
			\bottomrule
		\end{tabular}
	\end{center}
\end{table}

\begin{table}[h]
	\begin{center}
		\caption{Results of McNemar Test and Margin Test with Sample Disturbance for the Airway Hypertension Data, $\alpha = 0.05$.}
		\label{tab:21_SCT_sample_disturbance}
		\begin{tabular}{llccccc}
			\toprule			
			&		&	$n$	&	$n_{10}$	&	$n_{01}$	&	p-value	&	Decision	\\
			\hline
			Original Data	&	McNemar	&	21	&	7	&	1	&	0.0339	&	$H_{1}$	\\
			&	Margin	&	21	&	7	&	1	&	0.1033	&	$H_{0}$	\\
			Add One	&	McNemar	&	22	&	7	&	2	&	0.0596	&	$H_{0}$	\\
			&	Margin	&	22	&	7	&	2	&	0.2057	&	$H_{0}$	\\
			Reduce One	&	McNemar	&	20	&	6	&	1	&	0.0588	&	$H_{0}$	\\
			&	Margin	&	20	&	6	&	1	&	0.1588	&	$H_{0}$	\\				
			Adjust One	&	McNemar	&	21	&	6	&	2	&	0.1573	&	$H_{0}$	\\
			&	Margin	&	21	&	6	&	2	&	0.2957	&	$H_{0}$	\\	
			\bottomrule
		\end{tabular}
	\end{center}
\end{table}

The results of the equivalence test are presented in Table \ref{tab:21_SCT_sample_disturbance}, where the McNemar test rejects the null hypothesis under the significant level of $\alpha = 0.05$, which means that there is a significant difference between the probabilities of positive response before and after SCT. However, the margin test accepts the null hypothesis with the same level of significance. We propose the following sample disturbance method to  decide whether  to accept the null hypothesis or not, assuming that $n_{01} < n_{10}$ and there is a unit sample disturbance in favor of $H_{0}$:
\begin{enumerate}
	\item[(1)] Add one sample, which increases the sample size by one unit, adding a new observation such that $n = n+1, n_{10} = n_{10}, n_{01} = n_{01 } + 1$;
	\item[(2)] Reduce one sample, which removes an existing observation that is unfavorable to $H_0$ such that $n = n - 1, n_{10} = n_{10} -1, n_{01} = n_{01 }$;
	\item[(3)] Adjust one sample, which maintains the sample size and reduces the difference between the two discordant observations $n = n, n_{10} = n_{10} - 1, n_{01} = n_{01} + 1$.
\end{enumerate}

Since reducing the difference between $n_{10}$ and $n_{01}$ is a change in favor of $H_{0}$, the above three adjustment schemes make the new points move towards the acceptance region. Table \ref{tab:21_SCT_sample_disturbance} gives the results of the two tests with the four kinds of data, where the McNemar test rejects the null hypothesis and the margin test accepts $H_0$ with the original data. However, both tests accept the null hypothesis with the three adjusted observations, which means that the original data is on the boundary of the rejection region of the McNemar test, and the most closed three points to the original data fall into the acceptance region of the McNemar test. On the other hand, the original and deduced data are all in the acceptance region of the margin test. The difference between two discordant observations is not so significant, such that one unit of sample disturbance can lead to a different decision of the McNemar test. Based on the original data, we thus recommend accepting the null hypothesis $H_{0}$ as the difference between the two discordant observations is not sufficiently significant. To make a more accurate judgment, it is also necessary to increase the sample size to keep the observed data far from the boundary of the rejection region.

\begin{table}
	\begin{center}
		\caption{Frequency of Successfully Opening Box B by Male and Female Chimpanzees.}
		\label{tab:open_box}
		\begin{tabular}{ccccc}
			\toprule	
			&  &\multicolumn{2}{c}{Gender} &  \\
			\cline{3-4}
			&  & Male   & Fmale & SUM \\
			\hline
			& NO    & 12  & 9     &21 \\
			Open Box B	& YES               &27  & 17     & 44 \\		   
			& SUM              &39 & 26   & 65 \\		
			\bottomrule
		\end{tabular}
	\end{center}
\end{table}

\subsection{Box Opening and Gender}

\cite{smith_ruxton_2020} presented a behavioral ecology example, where $39$ male and $26$ female chimpanzees participated in the experiment to see if they could successfully open box B to get food among two puzzle boxes.
Table \ref{tab:open_box} presents the number of male and female chimpanzees who open box B.

\begin{table}[h]
	\begin{center}
		\caption{Results of McNemar test and Margin test with Sample Disturbance of Chimpanzee Data, $\alpha = 0.05$.}
		\label{tab:monkeys_sample_disturbance}
		\begin{tabular}{llccccc}
			\toprule			
			&		&	$n$	&	$n_{10}$	&	$n_{01}$	&	p-value	& Decision	\\
			\hline
			Original Data	&	McNemar	&	65	&	27	&	9	&	0.0027	&	$H_{1}$	\\
			&	Margin	&	65	&	27	&	9	&	0.0120	&	$H_{1}$	\\
			Add One	&	McNemar	&	66	&	27	&	10	&	0.0052	&	$H_{1}$	\\
			&	Margin	&	66	&	27	&	10	&	0.0193	&	$H_{1}$	\\
			Reduce One	&	McNemar	&	64	&	26	&	9	&	0.0041	&	$H_{1}$	\\
			&	Margin	&	64	&	26	&	9	&	0.0165	&	$H_{1}$	\\			
			Adjust One	&	McNemar	&	65	&	26	&	10	&	0.0077	&	$H_{1}$	\\
			&	Margin	&	65	&	26	&	10	&	0.0262	&	$H_{0}$	\\
			\bottomrule
		\end{tabular}
	\end{center}
\end{table}

Table \ref{tab:monkeys_sample_disturbance} presents the results of the McNemar test and the margin test with four kinds of data, where the results are the same for the original data, the added data, and the reduced data. And a different result of the two tests is from the adjusted data. It shows that the adjusted sample point $(26,10)$ falls in the acceptance region of the margin test, while the other three points $(27,9)$, $(27,10)$, and $(26,9)$ are all in the rejection regions of the two tests. Therefore the original data $(27,9)$ is actually in the rejection region of both tests, and we can reject the null hypothesis and accept that there is a significant correlation between the ability to open box B and the gender of chimpanzees.


\section{Discussion and Conclusion}

The bivariate binary distribution is a statistical model frequently used in biomedical statistics. Four joint probabilities $(p_{00}, p_{01}, p_{10}, p_{11})$ determine the structure of the model. And the correlation coefficient $\rho$ and two positive probabilities $p_{1+}, p_{+1}$ can describe the same model from the viewpoint of the marginal distribution. When the equivalence assumption holds, the two marginal probabilities are equivalent, such that the correlation coefficient $\rho$ and the positive probability $\pi$ can also determine the structure of the model.

According to the relationship between the joint and the marginal distribution, we determine the upper and lower bounds of the correlation coefficient $\rho$ and find the range of the marginal probability $\pi$ for a fixed correlation coefficient. 
Based on the two-variable parameter space of the null hypothesis, we discuss the symmetry of the marginal probability $\pi$ and the discordance probability $p_{10}$.
Using the conditional expectation method, we give the monotonicity of the cumulative distribution function of the discordant observation. And the minimum value of the distribution function is also determined by the correlation coefficient.
Using the joint distribution of two discordant observations, we propose an estimation algorithm for the confidence region of the two discordance observations. And an alternative equivalence test is derived from the resulting confidence region, called the margin test.

The McNear test is frequently used to test the equivalence hypothesis of the bivariate binary model, which assumes that the two dimensions are independent. 
To investigate the influence of the correlation between the two dimensions in the bivariate binary model on the result of the equivalence test,  we compare the acceptance region, size, and power of the McNear test and the margin test with different sample sizes and significance levels. 
The shape of the two acceptance regions is very similar, and the acceptance region of the margin test is slightly larger than the region of the McNemar test for the same sample size and significant level.
The range that the two tests correctly accept the null hypothesis gradually narrows with the increase in sample size. And the region that the margin test correctly identifies the parameters belonging to the parameter space of the null hypothesis is significantly larger than the corresponding region of the McNemar test for the same sample size, which means that the risk of type I error of the margin test is significantly smaller than the risk of the McNemar test.     
And it is found that the risk of type I error of the McNemar test is not monotonically decreasing with the increase of the correlation coefficient $\rho$ for relatively large sample size, while the type I error of the margin test does monotonically decrease with the increase of the correlation coefficient for all of the sample sizes.
From the contours of the power surface, we can find that the margin test tends to accept the null hypothesis when the discordant probability is relatively small, while the McNemar test tends to reject the null hypothesis with the same parameters.

There are two real-world examples in the previous section, one of the observed points lies on the boundary of the rejection region of the McNemar test, and the other point lies on the edge of the rejection region of the margin test. 
A unit sample disturbance method determines whether to accept the null hypothesis when the sample point lies between the boundaries of the two rejection regions.

We use the joint distribution of two discordant observations to construct the confidence region. The correlation coefficient is not directly used in the test process, so the influence of the correlation coefficient on the result of the equivalence test can only be observed when the null hypothesis holds. It is necessary to develop new statistical techniques to handle the observations with relatively small discordant probability or negatively correlated coefficients.









%
%


\bibliographystyle{plain}

\begin{thebibliography}{999}
	
	
	\bibitem{liu_ma_wu_tai_2006} J.P. Liu, M.C. Ma, C.Y. Wu, J.Y. Tai (2006) Tests of equivalence and non-inferiority for diagnostic accuracy based on the paired areas under
	ROC curves, Statist. Med., 25: 1219-1238.
	
	
	\bibitem{guo_luh_2020} J.H. Guo, W.M. Luh (2022) Testing two variances for superiority/non-inferiority and equivalence: using the exhaustion algorithm for sample size allocation with cost, British Journal of Mathematical and Statistical Psychology, 73, 316-332.
	
	
	\bibitem{sandie_molinari_etal_2022} A.B. Sandie, N. Molinari, A. Wanjoya, C. Kouanfack, 
	C. Laurent, J.B. Tchatchueng-Mbougua (2022) Non-inferiority test for a continuous variable with a flexible margin in an active controlled trial: an application to the "Stratall ANRS 12110/ESTHER" trail, Trials, 23: 202.
	
	
	\bibitem{paul_tiwari_etal_2021} E. Paul, R.C. Tiwari, S. Chowdhury, S. Ghosh (2021) A more powerful test for three-arm non-inferiority via risk difference: frequentist and Bayesian approaches, Journal of Applied Statistics, DOI: 10.1080/02664763.2021.1998391.
	
	
	\bibitem{ghosh_guo_ghosh_2022} S. Ghosh, W. Guo, S. Ghosh (2022) A hierarchical testing procedure for three arm non-inferiority trials, Computational Statistics and Data Analysis, 174,107521.
	
	\bibitem{book_Bishop_2007} Y.M.M. Bishop, S.E. Fienberg, P.W. Holland (2007) Discrete Multivariate Analysis Theory and Practice, Springer.
	
	\bibitem{fagerland_lydersen_laake_2013_same_unite}M.W. Fagerland, S. Lydersen, P. Laake (2014) Recommended tests and confidence intervals for paired binomial proportions, Statist. Med., 33, 2850-2875.
	
	
	
	\bibitem{fay_lumbard_2020}M.P. Fay, K. Lumbard (2021) Confidence interval for difference in proportions for matched pairs compatible with exact McNemar's or sign tests, Statist. Med., 40: 1479-1159.
	
	
	
	\bibitem{wu_2018}Y. Wu (2018) Power calculation of adjusted McNemar's test based on clustered data of varying cluster size, Biometrical Journal, 60: 1190-1200.
	
	
	\bibitem{fagerland_lydersen_laak_2013_sct}M.W. Fagerland, S. Lydersen, P. Laake (2013)  The McNemar test for binary matched-pairs data: mid-p and asymptotic are better than exact conditional, BMC Medical Research Methodology,  13:91. 
	
	
	\bibitem{rohmel_2005} J. Rohmel (2005) Problems with existing procedures to calculate exact unconditional p-value for non-inferiority/superiority and confidence intervals for two binomials and how to resolve them, Biometrical Journal, 47(1): 37-47.
	
	
	\bibitem{kang_chung_ahn_2005} S. H. Kang, S.J. Chung, C.W. Ahn (2005) Exact tests for one sample correlated binary data, Biometrical Journal 47(2): 188-193.
	
	
	
	
	\bibitem{nam_kwon_2009}J.M. Nam, D. Kwon (2009) Non-inferiority tests for clustered matched-pair data, Statist. Med., 28: 1668-1679.
	
	
	
	
	\bibitem{C_T_Le_1984} C.T. Le (1984) A symmetric bivariate binomial distribution and its application to the analysis of clustered samples in medical research, Biom. J., 26:3, 289-294. 
	
	
	
	\bibitem{albert_ingram_1985} A. W. Marshall, I. Olkin (1985) A family of bivariate distributions generated by the bivariate Bernoulli distribution,
	Journal of the American Statistical Association, 80:390, 332-338. 
	
	
	
	\bibitem{manuel_weib_silva_2014} M.G. Scotto, C.H. Weib,M.E. Silva, I. Pereira (2014) Bivariate binomial autogressive models, Journal of Multivariate Analysis, 
	125:233-251.
	
	\bibitem{wu_rai_yan_2019}X. Wu, J.P. Rai, G.Yan, Srivastava DK, S.N. Rai (2019) Correlation statistic for assessing a diagnostic value based on a bivariate Bernoulli distribution with application to retrospective single-center clinical study, International Journal of Clinical Biostatistics and Biometrics, 5:019.
	
	
	
	\bibitem{sidik_2003} K. Sidik (2003) Exact unconditional tests for testing non-inferiority in matched-pairs design, Statist. Med., 22: 265-278.
	
	
	
	
	
	
	\bibitem{lloyd_moldovan_2011} C.J. Lloyd, M.V. Moldovan (2011) Constructing more powerful exact tests of equivalence from binary matched pairs, Aust. N. Z. J. Stat., 53(1): 27-42.
	
	
	
	\bibitem{li_liu_goldberg_2011}X.C. Li, M.L. Liu, J.D. Goldberg (2011) A note on monotonicity assumptions for exact unconditional tests in binary matched-pairs designs, Biometrics, 67: 1666-1668.
	
	
	
	
	
	\bibitem{liu_hsueh_hsieh_chen_2002}J.P. Liu, H.M. Hsueh, E. Hsieh, J.J. Chen (2002) Tests for equivalence or non-inferiority for paired binary data, Statist. Med., 21: 231-245.
	
	
	
	
	
	\bibitem{fagerland_lydersen_laake_2014}M.W. Fagerland, S. Lydersen, P. Laake (2014) Recommended tests and confidence intervals for paired binomial proportions, Statist. Med., 33: 2850-2875.
	
	
	
	
	
	\bibitem{mohammadi_atashin_hofman_tan_2018} M. Mohammadi, A.A. Atashin, W. Hofman, Y. Tan (2018) Comparison of ontology alignment systems across single matching task via the McNemar's test, ACM Transactions on Knowledge Discovery from Data, 12(4), Article 51. 
	
	
	
	
	\bibitem{murtza_khan_akhtar_2019}I. Murtaza, A. Khan, N. Akhtar (2019) Object detection using hybridization of static and dynamic feature spaces and its exploitation by ensemble classification, Neural Comput \& Applic, 31: 347-361. 
	
	
	
	
	\bibitem{McNemar_test_1947}Q. McNemar (1947) Note on the sampling error of the difference between correlated proportions or percentages, Psychometrika., 12 (2): 153-157. doi:10.1007/BF02295996. PMID 20254758.
	
	
	\bibitem{bentur_lapidot_livnat_2009_SCT_example} L. Bentur, M. Lapidot, G. Livnat, F. Hakim, C. Lidroneta-Katz, I. Porat, D. Vilozni, R. Elhasid (2009) Airway reactivity in children before and after stem cell transplantation, Pediatr Pulm, 44:845-850.
	
	
	
	\bibitem{smith_ruxton_2020} M.P.Smith, G.D. Ruxton (2020) Effective use of the McNemar test, Behavioral Ecology and Sociobiology, 74: 133.
	
\end{thebibliography}

\end{document}